\documentclass[prb,aps,showpacs,twocolumn]{revtex4}
\usepackage{amssymb}
\usepackage{amsmath}
\usepackage{epsfig}
\usepackage{graphicx}

\begin{document}


\title{Fabrication and superconductivity of $Na_xTaS_2$ crystals}
\author{L. Fang$^{1,2}$, Y. Wang$^1$, P. Y. Zou$^1$, L. Tang$^1$, Z. Xu$^2$, H. Chen$^1$, C. Dong$^1$, L. Shan$^1$}
\author{H. H. Wen$^1$}\email{hhwen@aphy.iphy.ac.cn}

\affiliation{$^1$National Laboratory for Superconductivity,
Institute of Physics, Chinese Academy of Sciences, P.~O.~Box 603,
Beijing 100080, P.~R.~China.\\ $^2$Department of Material Science
and Engineering, Tongji University, Shanghai 200092, P.~R.~China}

\begin{abstract}
In this paper we report the growth and superconductivity of
$Na_xTaS_2$ crystals. The structural data deduced from X-ray
diffraction pattern shows that the sample has the same structure
as $2H-TaS_2$. A series of crystals with different superconducting
transition temperatures ($T_c$) ranging from 2.5 K to 4.4 K were
obtained. It is found that the $T_c$ rises with the increase of
$Na$ content determined by Energy-Dispersive x-ray
microanalysis(EDX) of Scanning Electron Microscope (SEM) on these
crystals. Compared with the resistivity curve of un-intercalated
sample $2H-TaS_2$ ($T_c$ = 0.8 K, $T_{CDW} \approx$ 70 K), no
signal of charge density wave (CDW) was observed in samples
$Na_{0.1}TaS_2$ and $Na_{0.05}TaS_2$. However, in some samples
with lower $T_c$, the CDW appears again at about 65 K. Comparison
between the anisotropic resistivity indicates that the anisotropy
becomes smaller in samples with more $Na$ intercalation (albeit a
weak semiconducting behavior along c-axis) and thus higher $T_c$.
It is thus concluded that there is a competition between the
superconductivity and the CDW. With the increase of sodium
content, the rise of $T_c$ in $Na_xTaS_2$ is caused mainly by the
suppression to the CDW in $2H-TaS_2$, and the conventional rigid
band model for layered dichalcogenide may be inadequate to explain
the changes induced by the slight intercalation of sodium in
$2H-TaS_2$.

\end{abstract}


\maketitle

\section{Introduction}
Layered transition-metal dichalcogenides (TMDC's) of the type
$MX_2$ ($M$ is the transition metal, X = S, Se, Te) have been
extensively studied for their rich electronic properties due to
low dimensionality. Each layer of TMDC's consists of a hexagonal
transition metal sheet sandwiched by two similar chalcogen sheets,
the interaction between the $MX_2$ layers is weak and van der
Waals-like. Charge density wave (CDW) and superconductivity (SC)
coexist in this kind of materials. The electron-phonon coupling
and its relationship with the CDW are investigated by angle
resolved photoemission in $2H-TaSe_2$ and $2H-NbSe_2$
systems\cite{Valla}. It is found that the CDW transition
temperature decreases and meanwhile the superconducting critical
temperature ($T_c$) increases from $TaSe_2$ through $TaS_2$ and
$NbSe_2$ to $NbS_2$, which indicates that these two orders (CDW
and superconductivity) compete each other\cite{1,2}.

Intercalation of atoms and molecules into the weak coupled region
between the $MX_2$ layers leads to significant modification of
properties, which is another highlight to study TMDC's. A variety
of atoms and molecules were reported to be intercalated into the
interlayer regions between the $MX_2$ layers, and the resulting
compounds are superconducting\cite{3}. Furthermore, different
superconducting transition temperatures $T_c$ were found depending
on the intercalated ions\cite{4,5}, and the $T_c$ increased when
the intercalated TMDC's were further hydrated\cite{6}. The change
of properties induced by intercalation may be explained in terms
of charge transfer from the intercalated atoms or molecules to the
host $MX_2$ layers, the band structure is unaltered upon
intercalation, and density of states (DOS) at Fermi surface (FS)
changes to reflect the transfer of charges from the intercalated
atoms or molecules. This picture is called as the rigid-band model
(RBM). The validity of this model is supported by some
photoemission experiments and calculations for various
intercalates of $MX_2$\cite{7,8,9,10}.

In recent years, however, some experiments and calculations
revealed that the change of properties of the intercalated $MX_2$
cannot be understood based on the rigid-band model. The electrical
measurement of $TaS_2$(pyridine)$_{1/2}$ showed that there was no
signal of CDW in the resistivity curve, which was attributed to
the suppression of a structural instability by
intercalation\cite{11}. Angle-resolved photoemission was used to
study the electronic band structures before and after the alkaline
metal $Cs$ and $Na$ were intercalated into the $2H-TaSe_2$ and
$VSe_2$\cite{12,13,14}, and it was found that the changes induced
by intercalation was more extensive than that expected by the
rigid-band model. The experiment of $Na$ intercalation into
$1T-TaS_2$ also revealed that the properties of the CDW phases and
the phase transitions are not only influenced by the shift of the
Fermi-level in the $Ta$ 5d band, but also by the superstructure of
the doped $Na$ in the interlayer region\cite{15}.

Therefore, it remains controversial whether it is possible to use
the rigid-band model to explain the changes induced by the
intercalation in TMDC's. In order to unravel this puzzle, more
efforts are desired. In the family of TMDC's, the sample $TaS_2$
may be one of the model systems to tackle this problem. Two basic
structures of $TaS_2$ were found and defined by the different
orientation of stacking chalcogen sheets, one is $1T-TaS_2$ with
$Ta$ in octahedral coordination with $S$ atoms, another is
$2H-TaS_2$ with $Ta$ in trigonal-prismatic coordination with $S$
atoms\cite{16,17}. The system $1T-TaS_2$ shows four CDW phase
transitions accompanied by changes in lattice parameters and
resistivity when temperature decreases, the CDW formation could be
explained by the Fermi surface nesting\cite{15,16,18}. However,
the system $2H-TaS_2$ is known for the existence of CDW order
below 70 K and a superconducting transition at 0.8 K\cite{16,18}.
When sodium ions are intercalated into $2H-TaS_2$, $T_c$
increases. Poly-crystalline samples of $Na_{0.33}TaS_2$ were made
and reported to be superconductors below 4.7 K
($T_c^{oneset}$)\cite{4,5}, and the $T_c$ was increased to 5.5 K
when the samples were hydrated\cite{6}. In this paper, we report
the growth of a series of crystals of $Na_xTaS_2$ (2H) with
different superconducting transition temperatures. Measurements of
X-ray diffraction and superconducting transition revealed good
quality of these crystals.

\section{Experiment}
Most layered compounds are known to intercalate atoms or molecules
into the inter-layer regions. A traditional way for fabricating
intercalated layered dichalcogend compounds consists of two steps,
the first step is to prepare poly-crystalline compound or single
crystal of $MX_2$ by chemical reaction or iodine-transport
reaction\cite{6,19}; Secondly, the intercalation is carried out by
chemical treatments, the samples are dipped into concentrated
solutions containing the intercalated atoms\cite{6,15,19}, or
exposed to metal vapors in a closed ampoule\cite{15}. A new method
suited for preparing samples for the studies of surface science
was also reported, the intercalated compounds were  prepared by
\emph{in situ} evaporation of metal atoms to the (0 0 0 1) cleaved
surface of the TMDC's crystals. This process should be realized
under the condition of ultra high vacuum (UHV)\cite{13,20}.

In this paper, we report the growth of crystals of $Na_xTaS_2$
directly with chemical reaction. The starting materials are
$NaCl$, $Ta$ powder and sulfur powder with purity higher than
$99.9${\%}. The nominal mole ratio of $NaCl$, $Ta$, and $S$ were
0.15, 1 and 2, respectively. They were mixed and thoroughly
ground, then pressed into pellet, finally the pellet was sealed in
an evacuated quartz tube. The tube was heated slowly up to 920
$^\circ C$ and  sintered for 48 hours, then slowly cooled down to
700 $^\circ C$ with a cooling rate of 1$\sim$2 $^\circ C/$hr.
Finally the quartz tube was cooled down with furnace by shutting
off the power. It is found that two kinds of crystals were
obtained. A large number of black thin platelets grew on the inner
side of the quartz tube. The surfaces of these platelets were
rather wrinkled, but the subsequent AC diamagnetism measurement
showed that these crystals had weak diamagnetic signal and broad
transition. Another type of crystals were also black thin
platelets which grew vertically on the surface of the pellet. The
surface of the second kind of crystals was shiny and mirror-like,
but the number of them is quite limited. Subsequent X-ray
diffraction pattern and AC diamagnetic measurement showed that
this kind of crystal has the typical structure of $2H-TaS_2$
single crystal and sharp superconducting transition is observed on
them. It is found that the size and the superconducting transition
temperature of the second type of crystals are related to the
cooling rate.

Fig. 1 shows an enlarged view of one piece of $Na_{0.1}TaS_2$
crystals. The crystal is shiny and thin platelet-like with clean
and smooth surface. The dimension of the crystal is about
1.7$\times$1 mm$^2$ and the thickness is about 0.1$mm$. Fig. 2
shows an enlarged view of the $Na_{0.1}TaS_2$ crystal at a corner.
It is clear that the sample has a layered structure.

\begin{figure}
\centering
\includegraphics[width=8cm]{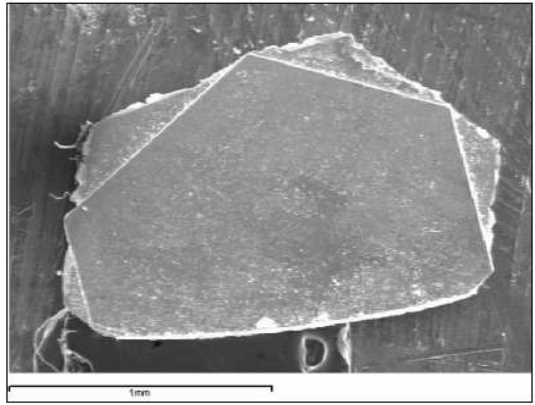}
\caption{An enlarged view of the $Na_{0.1}TaS_2$ crystal. The
intercalated sodium content is determined by Energy Dispersive
x-ray Microanalysis (EDX) of SEM.}\label{figure1}
\end{figure}

\begin{figure}
\centering
\includegraphics[width=8cm]{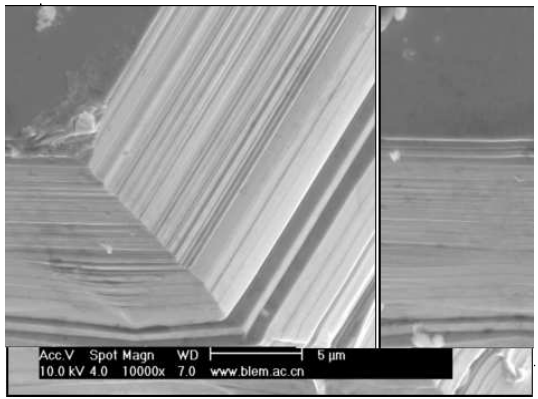}
\caption{An enlarged view of the $Na_{0.1}TaS_2$ crystal at a
corner. It is obvious that the samples have layered structure.}
\label{figure2}
\end{figure}

The x-ray diffraction pattern was performed at room temperature
employing a M18AHF x-ray diffractometer (MAC Science).
Crystallographic orientation and index are determined by Power-X,
a programm for processing X-ray diffraction data\cite{21}. The
magnetic and transport measurements were carried out with an
Oxford multi-parameter measurement system (Maglab-Exa-12). The
in-plane resistivity was determined by using standard four point
measurement. The off-plane resistivity was measured by putting two
separate pads of silver paste (one larger one for current and
smaller one for voltage) on both sides of the sample. Since the
in-plane resistivity is much smaller than the off-plane one, the
whole area of the sample plane was taken as the cross-section area
of the current flowing along c-aixs in determining $\rho_c$. The
microscopic and concentration analysis is achieved with Energy
Dispersive x-ray Microanalysis (EDX) of scanning electron
microscope(Oxford).

\section{ Results and Discussion}
\subsection{Structure and X-ray diffraction}
Fig.3 presents the X-ray diffraction pattern at room temperature
for one $Na_{x}TaS_2$ crystal. The XRD pattern is indexed on the
basis of hexagonal with $c= 12.082 \AA$, the value is quite close
to the reported value $c=12.070\AA$\cite{22} and
$c=12.097\AA$\cite{17}of $2H-TaS_2$.  Further comparison with the
value of $Na_{0.8}TaS_2$ ($c=14.36 \AA$) indicates that only
little content of sodium ions are intercalated into the
inter-layer regions of $2H-TaS_2$ in our present samples. The
space group is P\={6}m2, which is the subgroup of $2H-TaS_2$
(P6$_3$/mmc). For more $Na$ intercalated samples, the main peaks
remain unshifted and sharp, no obvious change of the $c-axis$
constant has been observed, but some commensurate modulation peaks
around each main peak along c-axis appear. Detailed analysis about
these modulations will be presented in a forthcoming paper.
Although the diffraction pattern is quite clean and the full-width
at the half maximum (FWHM) of the diffraction peak is narrow, we
cannot, however, rule out the possibility of some kind of stacking
fault which often appear in the intercalated crystals. However we
would argue that the stacking fault could be weak in our present
case since only very little sodium are intercalated into the
structure. Furthermore it remains unknown whether the intercalated
sodium atoms are in disordered state or organize into an ordered
lattice.

\begin{figure}
\centering
\includegraphics[width=9cm]{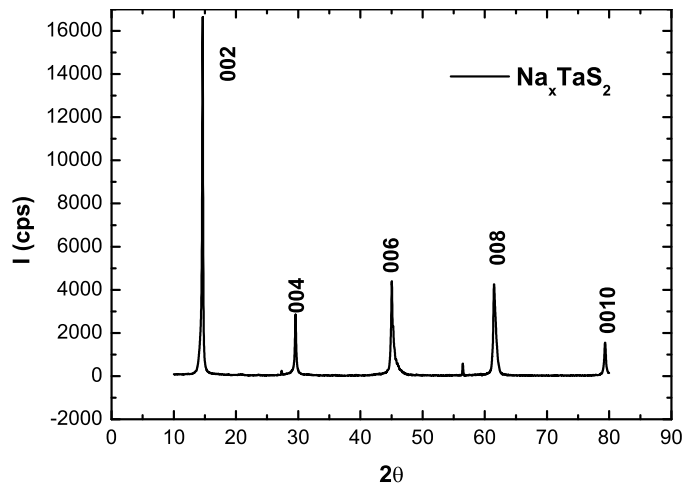}
\caption{X-ray diffraction pattern of a $Na_{x}TaS_2$ crystal. The
c-axis lattice constant determined here is about $c=12.082\AA$
which is quite close to that of $2H-TaS_2$. With increasing the
$Na$ content no obvious change of the $c-axis$ constant has been
observed in our samples.} \label{figure3}
\end{figure}

\begin{figure}
\centering
\includegraphics[width=9cm]{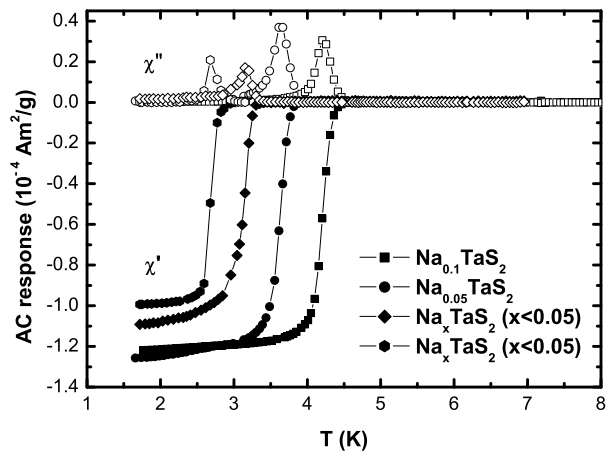}
\caption{ AC diamagnetism ($H_{AC}=1 Oe, f=333 Hz$) for a series
of $Na_xTaS_2$ crystals. Analysis using EDX reveals that the
sodium content are slight different among the samples.}
\label{figure4}
\end{figure}

\subsection{ AC diamagnetism and EDX of SEM}
It is found that both the $T_c$ and size of crystal are related to
the cooling rate. Fig. 4 shows four superconducting transition
curves of four samples. The $T_c$ determined in the diamagnetic
measurement changes from 4.4 K to 3.0 K, which is defined by the
onset point of the real part of AC susceptibility. The transition
width of each curve is less than 0.5 K, which indicates a complete
Meissner shielding effect. Because the biggest crystal has the
highest $T_c$, it is natural to suppose that the increase of $T_c$
is related to the change of sodium content, which is controlled by
the time and extent of diffusion. In order to know the sodium
content, these crystals were analyzed with the EDX of SEM. Taking
account of the possible inhomogeneity existing in the samples, we
selected multiple points on each crystal surface and analyzed. The
EDX spectrum showed that the atomic concentration of different
analyzed regions were similar, which indicates that these crystals
are homogeneous. Table-I gives sodium content of crystals analyzed
by EDX of SEM. $T_c$ rises roughly with the increase of $Na$
content in these crystals. As to the crystals with $T_c$ equal to
3.4 K and 3.0 K, respectively, little sodium ions are detected by
EDX. Because the $T_c$ of $2H-TaS_2$ is 0.8 K, thus, the sodium
content in these two crystals are approximately between 0 and
0.05.

\begin{table}
    \caption{Sodium extent intercalated into 2H-TaS$_2$ analyzed by EDX of SEM.}
    \begin{center}
        \begin{tabular}{|c|c|c|}
            \hline
            T$\mathrm{c}$ & Atomic {\%} (Na:Ta)& Formula \\
            \hline
            4.4 K &2.62 : 27.09 & Na$_{0.1}$TaS$_2$\\
            4.0 K& 1.22 : 25.90& Na$_{0.05}$TaS$_2$  \\
            3.4 K & sodium{\%} $ < 1$&Na$_x$TaS$_2$ (0$<$x$<$0.05) \\
            3.0 K& sodium{\%} $ < 1$&Na$_x$TaS$_2$  (0$<$x$<$0.05) \\
            \hline
        \end{tabular}
    \end{center}
\end{table}

\begin{figure}
\centering
\includegraphics[width=9cm]{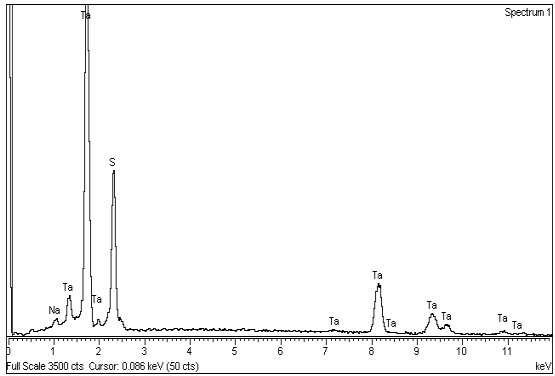}
\caption{ EDX spectrum of the $Na_{0.05}TaS_2$ crystal. The
characteristic peak of sodium around 1.1 keV is obvious.}
\label{figure5}
\end{figure}

Fig.5 shows the EDX spectrum of $Na_{0.05}TaS_2$ crystal. The
characteristic peak of sodium around 1.1 keV is obvious. No signal
of chlorine ions appears in the spectrum. The spectrum of
$Na_{0.1}TaS_2$ crystal is similar to that of $Na_{0.05}TaS_2$
crystal, but the peak of sodium gets enhanced for the former.

It is generally believed that superconductivity in the
dichalcogenides is due to electron-phonon coupling and is of
conventional BCS type. According to BCS theory, $T_c$ rises when
the density of states at Fermi surface is increased, which is
inferred by the following formula,

\begin{equation}
T_c\propto\Theta_{D}\exp[-1/VN(E_{F})]
 \label{eq1}
\end{equation}

where $\Theta_{D}$ is the Debye temperature, $V$ is the
electron-electron interaction which consists of the attractive
electron-phonon-induced interaction subtracted by the repulsive
Coulomb interaction, and $N(E_{F})$ is the density of states at
Fermi surface. As to our knowledge, change of $V$ was rarely
reported when atoms or molecules are intercalated into TMDC's.
Gample et al.\cite{3} pointed out that the electron-phonon
interaction is confined to the metallic disulfide layers and
insensitive to the distance between $MX_2$ layers caused by
intercalation. Thus, on the basis of little change of
electron-phonon interaction, it is considered that $N(E_{F})$
plays an important role for the increase of $T_c$ in $Na_xTaS_2$.

Calculations of DOS for the conduction band of $2H-TaS_2$ showed
that $E_F$ lies approximately at the mid-point of the $d_{z^2}$
band of $Ta$, and the $d_{z^2}$ band is half filled (occupied by
one electron per formula unit) and the DOS has a peak at the Fermi
level\cite{8,9}. When sodium ions are intercalated between the
layers of $2H-TaS_2$, charges are transferred from sodium to
$d_{z^2}$ band of $Ta$. The rigid-band model of intercalation
suggests that the host band structure is unaltered upon
intercalation, with $E_F$ increasing to reflect the transfer of
charges from the intercalant. Consequently the DOS at Fermi
surface should decrease. This is in contradiction to the
observation that the $T_c$ increases instead of decreasing when
the sodium atoms are intercalated. We will show below that the DOS
at Fermi level may increase because of the suppression to the CDW,
rather due to the charge transfer.

\subsection{The Anisotropy of the Upper Critical Field}
As to our knowledge, the upper critical field $H_{c2}(0)$ of
$Na_xTaS_2$ was rarely reported. We studied $H_{c2}(0)$ of one
piece crystal of $Na_{0.1}TaS_2$  with $T_c$ = 4.3 K through
diamagnetic and transport measurement. Fig.6 presents the AC
susceptibility under different magnetic fields. The transition
curve moves regularly down to lower temperatures with increasing
the magnetic field. Since the vortex motion is involved in the AC
susceptibility measurement, it is difficult to define the onset
point for superconductivity. Therefore in the following we will
use resistive transport measurement to determine the upper
critical field.

\begin{figure}
\centering
\includegraphics[width=9cm]{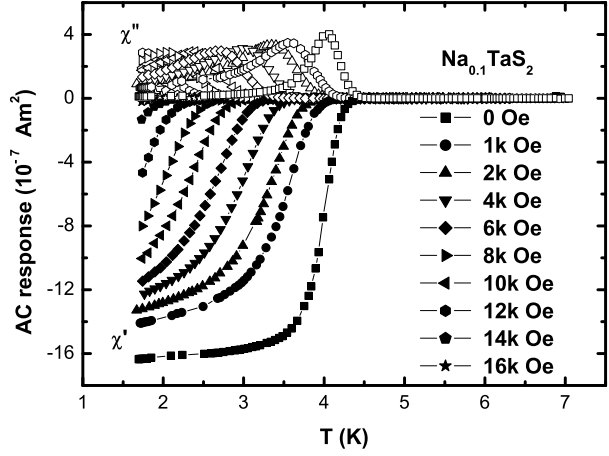}
\caption{ AC susceptibility ($H_{AC}=1 Oe, f=333 Hz$) under
different DC magnetic fields for the crystal Na$_{0.1}TaS_2$. Here
the DC magnetic field is applied along c-axis. } \label{figure6}
\end{figure}

Fig.7 shows the in-plane resistivity (ab plane) of the crystal
$Na_{0.1}TaS_2$ at zero field. Decreasing temperature from $10 K$
to $4.9 K$, the resistivity $\rho$ changes subtly, and an abrupt
superconducting transition happens at $4.83 K$, zero resistivity
is obtained at about $4.3 K$. A sharp transition with width less
than $0.5 K$ indicates the good quality of the crystal
$Na_{0.1}TaS_2$. The zero superconducting transition temperature
is $4.3 K$, which is consistent with the result of AC
susceptibility.

\begin{figure}
\centering
\includegraphics[width=9cm]{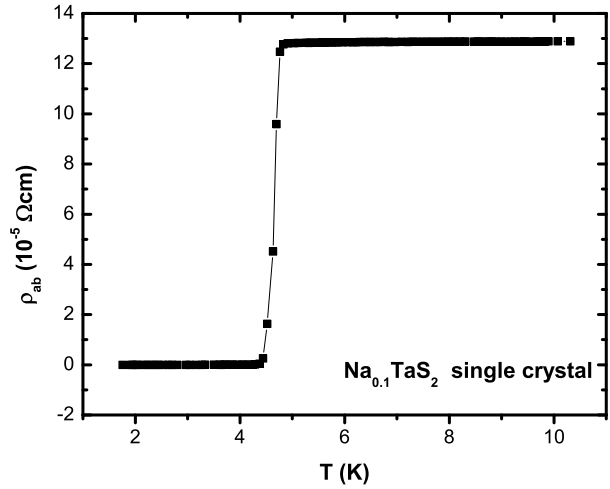}
\caption{ In-plane resistivity (ab plane) vs. temperature for one
piece of crystal $Na_{0.1}TaS_2$ under zero field. The transition
width is less than 0.5 K.} \label{figure7}
\end{figure}

\begin{figure}
\centering
\includegraphics[width=9cm]{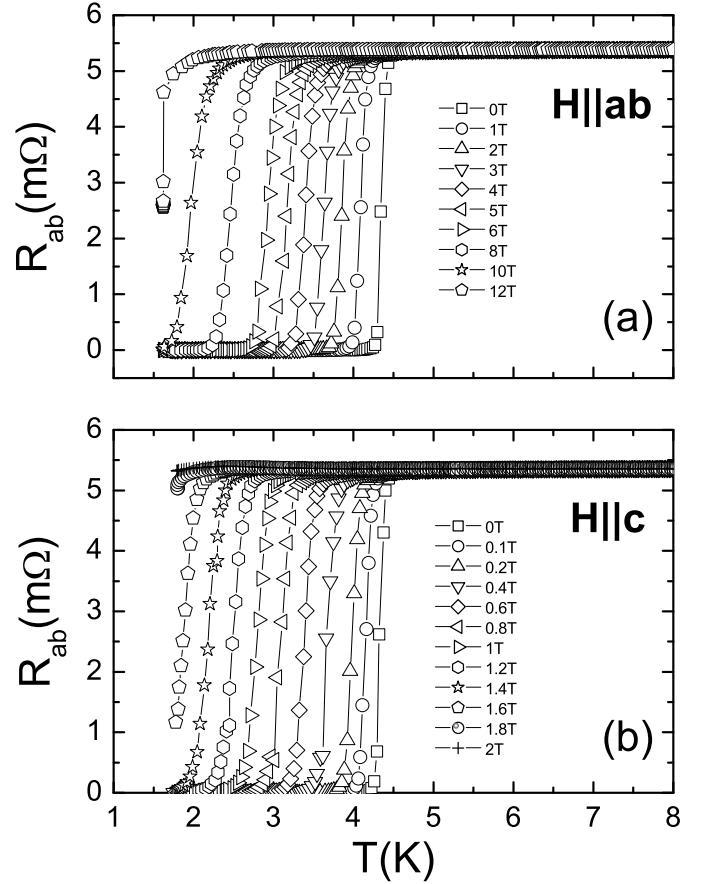}
\caption{ The superconducting transition measured at different
magnetic fields when the field is applied (a) perpendicular to and
(b) parallel to c-axis. From the mid-point of the resistive curve
one can determine the upper critical field $H_{c2}(T)$.}
\label{figure8}
\end{figure}

\begin{figure}
\centering
\includegraphics[width=9cm]{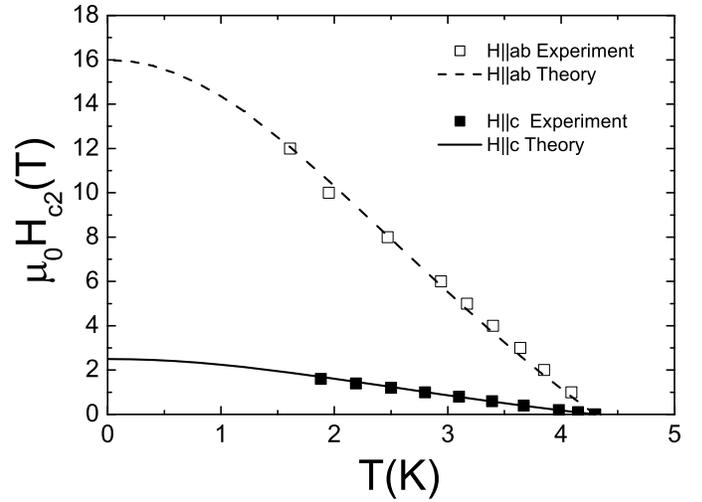}
\caption{The upper critical field determined from the mid-point of
the transition curve. The solid and dashed lines here are
theoretical curves of $H_{c2}(T)=H_{c2}(0)[(1-t^2)/(1+t^2)]$ with
$H_{c2}^{||c}(0)=2.5 T$ and $H_{c2}^{||ab}(0)=16 T$.}
\label{figure9}
\end{figure}

In order to determine the upper critical field and the anisotropy
of superconductivity, we measured the resistive transitions at
different magnetic fields with the filed direction along
$ab-plane$ (Fig.8(a)) and $c-axis$ (Fig.8(b)). One can see that
the resistive curve shifts parallel down to lower temperatures
with the increase of magnetic field. From the mid-point of the
transition curves, we determine the upper critical fields for both
directions which are shown in Fig.9. In the Ginzburg-Landau
theory, it is known that $H_{c2}=\Phi_0/2\pi\xi^2$ and $\xi\propto
\sqrt{(1+t^2)/(1-t^2)}$, with $\Phi_0$ the flux quanta, $\xi$ the
coherence length, $t=T/T_c$ the reduced temperature, thus one has

\begin{equation}
H_{c2}(T)=H_{c2}(0)\frac{1-t^2}{1+t^2}
\label{eq2}
\end{equation}

We use above equation to fit our data and show them as the solid
and dashed lines in Fig.9. The zero temperature upper critical
fields $H_{c2}(0)$ determined in this way are $H_{c2}^{||c}(0)=2.5
T$ and $H_{c2}^{||ab}(0)=16 T$, therefore the anisotropy
$H_{c2}^{||ab}(0)/H_{c2}^{||c}(0)$=$\xi_{ab}(0)/\xi_{c}(0)$=$\sqrt{m_{c}/m_{ab}}$=6.4,
 where $m_c$ and $m_{ab}$ are the effective mass tensors when the electrons are moving perpendicular and parallel to the $TaS_2$ layers. This value is quite close to that of optimally doped $YBa_2Cu_3O_7$.
This is to our surprise since the sample here is clearly of
thin-platelet shape which looks like the much more anisotropic
curate system $Bi_2Sr_2CaCu_2O_8$ single crystals.  Actually the
zero temperature value of upper critical field can also be
determined through the Werthamer-Helfand-Hohenberg (WHH)
formula\cite{23}:

\begin{equation}
H_{c2}(0)=-0.693T_c(\frac{dH_{c2}}{dT})_{T=T_c}
\label{eq3}
\end{equation}

Here $dH_{c2}/dT$ is the slope of $H_{c2}(T)$ near $T_c$, which is
about 0.7163 T/K for $H||c$ and 4.5 T/K for $H||ab$. Using above
formula the zero temperature values of upper critical fields are
$H_{c2}^{||c}(0)=2.13 T$ and $H_{c2}^{||ab}(0)=13.4 T$, which are
close to the values determined in fitting the data to eq.(2).

\subsection{Suppression to CDW by Sodium Intercalation}
Trigonal prismatic layer compound $2H-TaS_2$ generally exhibits a
charge-density-wave related phase transition accompanied by a drop
in resistivity around $70 K$\cite{11}. It is thus interesting to
measure the resistivity of our samples to high temperatures to see
whether the CDW transition is still there. We thus measured the
resistivity of crystals of $Na_{0.1}TaS_2$ from $2 K$ to $300 K$.
It is found that the resistivity $\rho$ decreases with the
temperature smoothly and the superconducting transition happens at
$4.4 K$, no sudden drop of resistance on the resistivity curve was
observed above $T_c$. Fig. 10 shows the comparison of resistivity
between undoped $2H-TaS_2$ and our sample $Na_{0.1}TaS_2$. From
here it is tempting to conclude that the CDW is completely
suppressed in our samples.

\begin{figure}
\centering
\includegraphics[width=9cm]{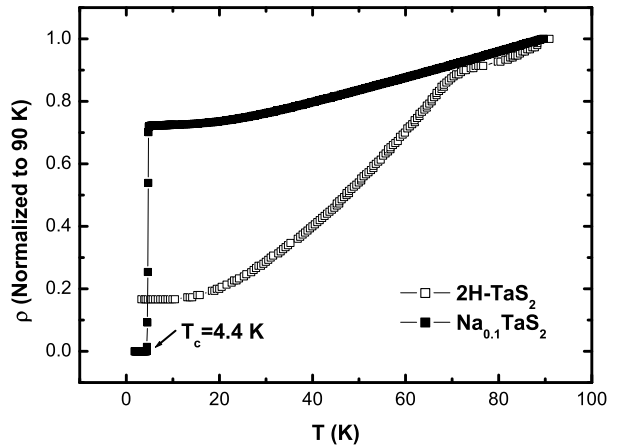}
\caption{Comparison of resistivity between $2H-TaS_2$ and
$Na_{0.1}TaS_2$. For clarity, here we show only the data from $2
K$ to $90 K$. Resistivity of $2H-TaS_2$ is adopted from Ref.11. }
\label{figure10}
\end{figure}

\begin{figure}
\centering
\includegraphics[width=9cm]{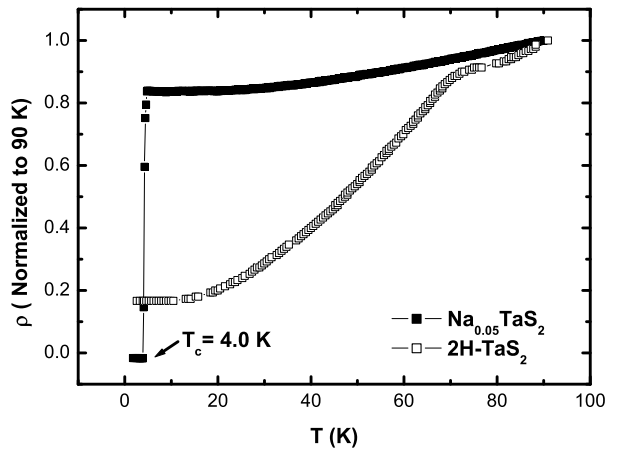}
\caption{ Comparison of  resistivity between $2H-TaS_2$ and
$Na_{0.05}TaS_2$. For clarity we show here only the data from $2
K$ to $90 K$. Resistivity of $2H-TaS_2$ is adopted from ref. 11.}
\label{figure11}
\end{figure}

For further verification of the result obtained from
$Na_{0.1}TaS_2$, we measured the resistivity of one crystal
$Na_{0.05}TaS_2$ and show them in Fig.11. The CDW-like drop in
resistivity curve is also completely absent in the temperature
region from $2 K$ to $300 K$. This result provides convincing
evidence that there is a competition between the superconductivity
and charge density wave in layered chalcogenide $Na_xTaS_2$. When
sodium ions are intercalated into $2H-TaS_2$, CDW order is
destroyed and $T_c$ increases. The suppression to the CDW may be
understood by the better c-axis conduction after the sodium
intercalation. In this case the system deviates from two
dimensionality as in $2H-TaS_2$ and thus prevents the lattice
instability. This is partially supported by the relatively small
anisotropy of $m_{c}/m_{ab}$ as determined above in the sodium
intercalated samples. When the CDW is suppressed, the effective
DOS at the Fermi surface is eventually enhanced leading to a much
higher $T_c$.

\begin{figure}
\centering
\includegraphics[width=9cm]{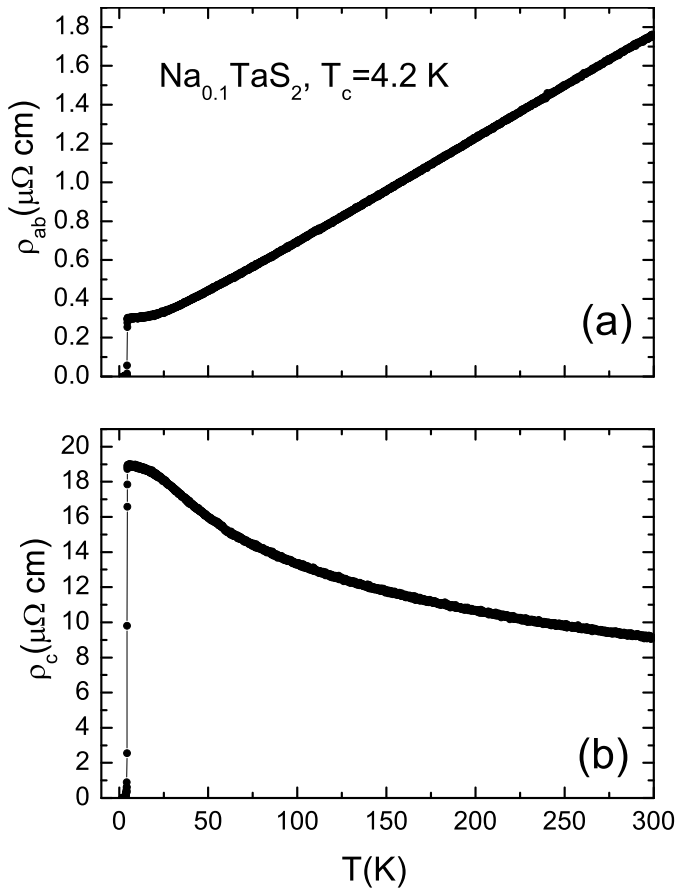}
\caption{Temperature dependence of (a) in-plane and (b) off-plane
resistivity for the sample $Na_{0.1}TaS_2$. One can see that the
in-plane resistivity shows a metallic behavior, but the off-plane
one has an insulating behavior above $T_c$. The anisotropy of
resistivity is about 63.6 at $10 K$ and 5.3 at $300 K$. On both
curves we cannot see any trace of CDW. A linear behavior of the
in-plane resistivity is observed in wide temperature range. This
resembles that in optimally doped high-$T_c$ cuprate
superconductors.} \label{figure12}
\end{figure}

\begin{figure}
\centering
\includegraphics[width=9cm]{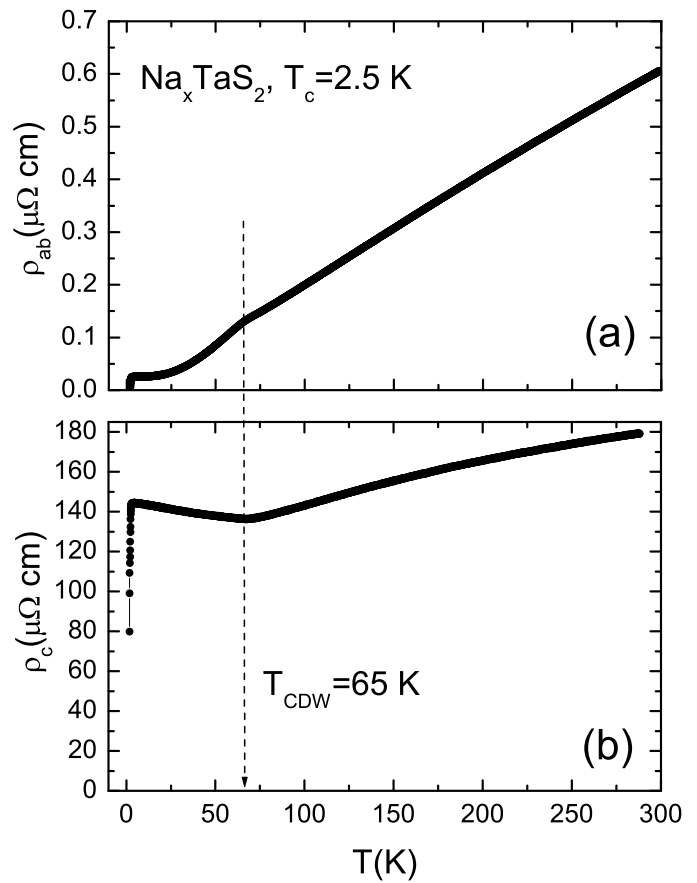}
\caption{Temperature dependence of (a) in-plane and (b) off-plane
resistivity of a sample $Na_xTaS_2$ with $x\leq 0.05$ and $T_c=2.5
K$. A clear kink corresponding to the CDW transition can be seen
here. One can also see that the in-plane resistivity shows
metallic behavior in wide region of temperature, but the off-plane
resistivity first shows a metallic behavior above $T_{CDW}$ and an
insulating behavior below $T_{CDW}=65 K$.  The anisotropy of
resistivity is about 5000 at $10 K$ and 300 at $300 K$. }
\label{figure13}
\end{figure}

\subsection{Anisotropy of Resistivity and the Competition between Superconductivity and CDW}
In pyridine intercalated systems, the anisotropy of DC resistivity
is in the order of $10^5$ along c-axis and ab-plane. This
extremely high anisotropy is far beyond the value we can expect in
our present system. Therefore intercalating sodium here may
enhance the electrical conduction along c-axis and suppress the
feature of two dimensionality. To check whether this is true, we
measured the in-plane and off-plane resistivity for two samples,
one with $x=0.10$ and $T_c=4.2 K$, and another with $x\leq 0.05$
and $T_c=2.5 K$ (determined as the mid-point of the resistive
transition). In Fig.12, we present the temperature dependence of
the (a) in-plane and (b) off-plane resistivity for the sample with
$x=0.10$ and $T_c=4.2 K$. No any trace of CDW can be observed
here. Furthermore, one can easily see that the temperature
dependence of the in-plane resistivity for this sample is rather
linear in quite wide temperature region. This result has been
repeated in all our samples with $T_c$ higher than 4 K. This
behavior resembles that of cuprate superconductors at optimal
doping (with the highest $T_c$ in the same system). It remains an
interesting question that whether this linear behavior observed
here has any intact correlation with that in high-$T_c$ cuprate
superconductors. We can also determine the anisotropy of
resistivity for this sample at different temperatures. It is found
that $\rho_c/\rho_{ab}$ is about 63.6 at $10 K$ and 5.3 at $300
K$. Surprisingly the off-plane resistivity shows a clear
semiconducting behavior indicating totally different electron
conduction behavior when current is flowing along the plane or
perpendicular to it. This interesting weak semiconducting behavior
of c-axis resistivity has also been repeated in our samples with
high $T_c$. Accompanying with this semiconducting behavior, a
c-axis lattice modulation (with the modulation spacing of about
$4c$) has been observed only in these samples with high $T_c$.
Detailed analysis about the correlation with the resistive
transport property, CDW and superconductivity is under way and
will be published separately. For the less intercalated sample
($x\leq 0.05$ and $T_c=2.5 K$), the situation becomes very
different. As shown in Fig.13(a) and (b), one can see that a CDW
transition occurs at about $65 K$. Worthy of noting here is the
much higher anisotropy of the resistivity. One can see that
$\rho_c/\rho_{ab}$ is about 5000 at $10 K$ and 300 at $300 K$. The
anisotropy of resistivity in this less intercalated sample is
about 60 to 100 times higher than that in the more intercalated
sample ($x=0.1, T_c=4.2 K$) although a metallic behavior is
observed above $T_{CDW}$ in these less intercalated samples. The
much more higher anisotropy in these less intercalated samples may
also interpret why the CDW appears in this sample. The strange
metal-insulator (M-I) transition of the off-plane resistivity at
$T_{CDW}$ in the present sample is a very interesting issue and
the discussion on it may exceed the scope of this paper. It must
be mentioned that this M-I transition can only be observed in
relatively thick samples (above $0.2\mu m$). In rather thin
samples, the c-axis resistivity shows the similar behavior of the
in-plane resistivity without showing the M-I transition at the
$T_{CDW}$. We attribute the disappearance of the M-I transition to
the significant in-plane component of the total voltage drop on
the two voltage leads attached to the two opposite surfaces if the
leads are slightly asymmetric. Detailed analysis and discussion of
the temperature dependence of resistivity for samples with
systematic intercalated $Na$ content will be presented in a
forthcoming paper. We must point out that, since the thickness of
the sample is very thin (typically in the scale of $0.1 mm$ to
$0.2 mm$), to precisely measure the off-plane resistivity is
really a problem. Thus it may subject to corrections in the future
with more refined measurements. However the correction, if any in
the future, will be within the range of about $20\%$, especially
for the in-plane resistivity. In addition, we believe that the
general temperature dependence of resistivity measured here will
not be altered too much. Therefore the resistive data measured
here may provide an useful platform for further understanding the
interplay between CDW and superconductivity. Although we are not
sure why the slight intercalation of $Na$ can change the
anisotropy and the electronic behavior so drastically, it may be
safe to conclude that the higher anisotropy in the sample with
less $Na$ content enhances the 2D behavior and thus favors the
structural instability leading to the CDW transition. This
consequently suppresses the effective DOS at $E_F$ and lowers the
superconducting transition temperature.

\section{Conclusion}

A new way to grow crystals of $Na_xTaS_2$ is presented. A series
of crystals with different superconducting transition temperatures
$T_c$ ranging from 2.5 K to 4.4 K were obtained. It is found that
$T_c$ rises with the increase of $Na$ content determined from the
EDX of SEM. Compared with the resistivity curve of
$2H-TaS_2$($T_c$ = 0.8 K,$T_{CDW}\approx$ 70 K), no signal of
charge density wave (CDW) was observed in our present samples
$Na_{0.1}TaS_2$ and $Na_{0.05}TaS_2$. The upper critical field and
its anisotropy (about 6.4 for sample with $x=0.1$ and $T_c=4.2 K$)
have also been determined. The anisotropy of resistivity is
strongly suppressed together with the missing of the CDW in
samples with higher $T_c$ and more $Na$ content. In samples with
less $Na$ content and lower $T_c$, the CDW can be easily observed.
It is concluded that there is a competition between the
superconductivity and the CDW order: The rise of $T_c$ in
$Na_xTaS_2$ by increasing the sodium content may be caused by the
increase of DOS at Fermi surface when the CDW is suppressed.
{\bf Acknowledgements} This work is supported by the National
Science Foundation of China, the Ministry of Science and
Technology of China, and the Chinese Academy of Science within the
knowledge innovation project. The authors are grateful to C. H.
Wang, W. W. Huang and S. L. Jia for technical assistance and
helpful discussions about some measurements. The authors thank
Huan Yang for helping to transform the graph form.


\end{document}